\documentstyle{elsart}
\begin{document}
\begin{frontmatter}

\title{Quasilocal energy-momentum for \\ geometric gravity theories}

\thanks[jmn]{Email: nester@joule.phy.ncu.edu.tw}
\thanks[rst]{Email: m792001@joule.phy.ncu.edu.tw}

\author{Chiang-Mei Chen},
\author{James M. Nester\thanksref{jmn}}
and
\author{Roh-Suan Tung\thanksref{rst}}
\address{Department of Physics,
National Central University, Chung-Li, Taiwan 32054}

\begin{abstract}
{}From a covariant Hamiltonian formulation, using symplectic ideas, we
obtain covariant quasilocal energy-momentum boundary expressions
for general gravity theories.
The expressions depend upon which
variables are fixed on the boundary, a reference configuration
and a displacement vector field.
We consider applications to Einstein's theory, black hole
thermodynamics and alternate spinor expressions.
\end{abstract}

\end{frontmatter}

\section{Introduction}

The source of gravity is the energy-momentum density of all other
physical fields.  For the gravitational field itself, however,
energy-momentum is not so simply described.  Although geometries with
suitable asymptotic regions have a well defined {\em total\/}
energy-momentum, the equivalence principle precludes any proper
{\em local\/} density.  Hence, the idea of quasilocal quantities for
gravity has been advocated and there have been many recent proposals.
Here we present a set of {\em quasilocal\/} energy-momentum
expressions for quite general geometric gravity theories.
Distinguishing features of our Hamiltonian based expressions are that
they are covariant and that the different possible expressions are
associated with the choice of variables to be held fixed on the
boundary and a symplectic boundary variational structure.

\section{The covariant Hamiltonian}

We consider general theories of dynamic geometry.
The possible geometric potentials are the {\em coframe\/}
$\vartheta^\alpha$
and {\em connection\/} $\omega^\alpha{}_\beta$ one-forms and the
{\em metric\/} coefficients $g_{\mu\nu}$.
The corresponding field strengths are the {\em torsion\/}
$ \Theta^\alpha := D \vartheta^\alpha
 = d \vartheta^\alpha + \omega^\alpha{}_\beta \wedge \vartheta^\beta$
and {\em curvature\/}
$ \Omega^\alpha{}_\beta := d \omega^\alpha{}_\beta
  + \omega^\alpha{}_\gamma \wedge \omega^\gamma{}_\beta $
2-forms and the {\em nonmetricity\/} one-form
$ {\cal G}_{\mu\nu} := D g_{\mu\nu}
  = d g_{\mu\nu} - \omega^\alpha{}_\mu g_{\alpha\nu}
                 - \omega^\alpha{}_\nu g_{\mu\alpha} $.
In place of the usual ``second order'' Lagrangian 4-form
$ {\cal L} = {\cal L}(g,\vartheta,\omega,{\cal G},\Theta,\Omega) $,
we introduce covariant canonically conjugate momenta forms to obtain
a ``first order'' Lagrangian 4-form
\begin{equation}
{\cal L}
 :={\cal G}_{\mu\nu} \!\wedge\! \pi^{\mu\nu}
  + \Theta^\alpha \!\wedge\! \tau_\alpha
  + \Omega^\alpha{}_\beta \!\wedge\! \rho_\alpha{}^\beta
  - \Lambda( g, \vartheta; \pi, \tau, \rho),
\end{equation}
which yields ``first order'' field equations via independent
variations of the fields $g, \vartheta, \omega$ and momenta
$\pi, \tau, \rho$.

For any fixed slicing of space-time by
t=constant surfaces $\Sigma$ along with a connecting vector field $N$,
the decomposition of the first order Lagrangian 4-form according to
the general pattern $ {\cal L} \equiv dt \wedge i_N {\cal L} =
dt \wedge ({\hbox{\it \char'44}\!}_N \varphi\wedge p -{\cal H}(N) )$
identifies the {\em covariant Hamiltonian\/} 3-form (or density)
 \cite{Ne91}, i.e., the generator of evolution along $N$:
\begin{eqnarray}
{\cal H}(N)
 &=& i_N \Lambda
  + {\cal G}_{\mu\nu} \!\wedge\! i_N \pi^{\mu\nu}
  -  i_N \vartheta^\alpha D \tau_\alpha
  - \Theta^\alpha \!\wedge i_N\! \tau_\alpha
  -\Omega^\alpha{}_\beta \!\wedge\! i_N \rho_\alpha{}^\beta
  \nonumber \\
 &-& i_N \omega^\alpha{}_\beta ( D \rho_\alpha{}^\beta
  - g_{\alpha\nu} \pi^{\beta\nu}-g_{\mu\alpha}\pi^{\mu\beta}
 + \vartheta^\beta \wedge \tau_\alpha)
  +  d {\cal B}(N),
\end{eqnarray}
where
$ {\cal B}(N) := i_N \vartheta^\alpha \tau_\alpha
 + i_N \omega^\alpha{}_\beta \rho_\alpha{}^\beta $.

\section{Covariant quasilocal expressions}

Noether's theorem for a translation applied to ${\cal L}$
yields the differential identity $ d{\cal H}(N) \equiv $
(terms $\propto$ field equations) and the algebraic identity
$ {\cal H}(N) \equiv$ (terms $\propto$ field equations)
$+ d {\cal B}(N)$.  Consequently, the Hamiltonian
$\int_{\Sigma}{\cal H}(N)$,
the integral of the density over a finite spatial region, has a
conserved value on a solution given via Stokes theorem by
$\oint_{\partial \Sigma}{\cal B}(N)$.
This boundary integral gives the {\em quasilocal quantities\/} (energy
etc.);  the limiting value at infinity should be a total conserved
quantity for $N$ asymptotically Killing.

However, the value of the Hamiltonian is not yet firmly fixed.
As with other Noether currents we can add any exact
differential thereby modifying the
expression given above for the Hamiltonian
boundary term.  Moreover, such an adjustment only affects the
boundary term in the variation of the Hamiltonian and thus does
not affect the basic physics: the dynamic equations.
The Hamiltonian
formulation, however, includes a further principle which fixes the
possible forms of ${\cal B}$. The proper  form is identified by
considering the variation of the Hamiltonian density:
$\delta {\cal H}(N) = $ (field eq. terms)
$ + d i_N (\delta g_{\mu\nu} \pi^{\mu\nu}
  + \delta \vartheta^\alpha \wedge \tau_\alpha
  + \delta \omega^\alpha{}_\beta \wedge \rho_\alpha{}^\beta) $.
The total differential term in $\delta {\cal H}(N)$ produces a
boundary integral which reflects both the choice
of variables to be held fixed (``control'' variables) and the
{\em symplectic structure\/} \cite{KT79}. The boundary term in
$\delta {\cal H}(N)$  vanishes (as it should) for fixed control
variables on a finite boundary.  But for a boundary at infinity
the limit of the asymptotic fall offs generally gives a nonvanishing
results (e.g., for Einstein's theory) then the above ${\cal
B}(N)$ expression needs adjustment \cite{RT74}.

An acceptable boundary expression for general metric compatible
gravity theories was found \cite{Ne91} and soon
improved by Hecht \cite{He95}.  From his work we saw more
possibilities.  The expressions for the Hamiltonian
boundary term which are 4-{\em covariant\/} and yield a 4-{\em
covariant symplectic\/} structure for the boundary term in $\delta
{\cal H}(N)$ were identified \cite{CMC94}. There are two types:
\begin{equation}
 {\cal B}_\varphi(N) := i_N \varphi \wedge \Delta p
 - \varsigma \Delta \varphi \wedge i_N
 {\buildrel \scriptstyle \circ \over p},  \label{Bphi}
\end{equation}
\begin{equation}
{\cal B}_p(N) := i_N {\buildrel \scriptstyle \circ \over
\varphi} \wedge \Delta p - \varsigma\Delta\varphi\wedge i_N p,
 \label{Bp}
\end{equation}
depending upon whether the configuration field k-form $\varphi$ or
its conjugate momenta $p$ is ``controlled'' on the boundary
\cite{KT79}
(i.e., either Dirichlet or Neumann boundary conditions).
 Here ${\buildrel \scriptstyle \circ \over \varphi}$
and ${\buildrel \scriptstyle \circ \over p}$ are the values in a
reference configuration,
$\Delta\varphi=\varphi-{\buildrel\scriptstyle\circ \over \varphi}$,
$\Delta p = p - {\buildrel \scriptstyle \circ \over p}$
and $\varsigma = (-1)^k$.
Thus for the geometric fields
\begin{eqnarray}
{\cal B}(N) &=&
\left\{ \begin{array}{c}
 -\Delta g_{\mu\nu}
i_N{\buildrel\scriptstyle\circ\over\pi}{}^{\mu\nu}\\
 -\Delta g_{\mu\nu} i_N \pi^{\mu\nu}
 \end{array} \right\}
+
\left\{ \begin{array}{c}
 i_N \vartheta^\alpha \Delta \tau_\alpha
   + \Delta \vartheta^\alpha \wedge i_N
   {\buildrel \scriptstyle \circ \over \tau}_\alpha \\
 i_N {\buildrel \scriptstyle \circ \over \vartheta}{}^\alpha
   \Delta \tau_\alpha
   + \Delta \vartheta^\alpha \wedge i_N \tau_\alpha
 \end{array} \right\}  \nonumber \\
&+&
\left\{ \begin{array}{c}
 i_N \omega^\alpha{}_\beta \Delta \rho_\alpha{}^\beta
   + \Delta \omega^\alpha{}_\beta \wedge i_N
   {\buildrel \scriptstyle \circ \over \rho}{}_\alpha{}^\beta \\
 i_N {\buildrel \scriptstyle \circ \over \omega}{}^\alpha{}_\beta
   \Delta \rho_\alpha{}^\beta
   + \Delta \omega^\alpha{}_\beta
   \wedge i_N \rho_\alpha{}^\beta
 \end{array} \right\}\ ,
\end{eqnarray}
\noindent
where for each bracket the upper (lower) line is to be used if
the field (momentum) is controlled.
Hence, as in thermodynamics, there are several kinds of ``energy'',
each corresponds to the work done in a different (ideal) physical
process \cite{KT79}.
Our quasilocal boundary expressions are {\em covariant\/}---aside from
the manifestly non-covariant explicit connection terms in
${\cal B}$.  These terms include a real physical effect plus an
unphysical dynamical reference frame effect.  These effects can be
separated using the identity
\begin{equation}
(i_N \omega^\alpha{}_\beta) \vartheta^\beta \equiv
 i_N \Theta^\alpha + D N^\alpha
 - {\hbox{\it \char'44}\!}_N \vartheta^\alpha. \label{inomega}
\end{equation}
Via this identity the boundary term contains a time derivative of
certain frame components.  Such terms have been noted previously
in Einstein's theory \cite{Kij83}.

The variation of the Hamiltonian, in addition to the field equation
terms, now includes for each variable one of the boundary terms
\begin{equation}
d i_N (\delta \varphi \wedge \Delta p), \qquad \hbox{or} \qquad
d i_N (- \Delta \varphi \wedge \delta p), \label{symp}
\end{equation}
which reflect the {\em symplectic\/} structure and the control mode.
Specifically, for the geometric variables, the total differential
term in $\delta {\cal H}(N)$ is of the form $di_N {\cal C}$ where
\begin{equation}
{\cal C}=
 \left\{ \begin{array}{c}
   \delta g_{\mu\nu} \Delta \pi^{\mu\nu} \\
   - \Delta g_{\mu\nu} \delta \pi^{\mu\nu}
   \end{array} \right\}
+
 \left\{ \begin{array}{c}
   \delta \vartheta^\alpha \wedge \Delta \tau_\alpha \\
   - \Delta \vartheta^\alpha \wedge \delta \tau_\alpha
   \end{array} \right\}
+
 \left\{\begin{array}{c}
\delta \omega^\alpha{}_\beta \wedge \Delta \rho_\alpha{}^\beta \\
  - \Delta \omega^\alpha{}_\beta \wedge \delta \rho_\alpha{}^\beta
   \end{array} \right\} , \label{C}
\end{equation}
here again the upper (lower) line in each bracket corresponds to
controlling the field (momentum).  Our quasilocal expressions
are {\em uniquely\/} determined by the Hamiltonian variation (\ref{C})
and
the requirement that all of the quasilocal quantities {\em vanish\/}
when the fields have the reference configuration values.

This general formalism readily specializes to coordinate or
orthonormal frames and Riemannian, Riemann-Cartan or teleparallel
geometry. For General Relativity and the Poincar\'e Gauge Theory
(asymptotically flat {\em or\/} constant curvature)
our expressions reduce to known ones \cite{BO87,HS85}
and give the total quantities at spatial
{\em and\/} future null infinity \cite{HN95}.
Non-vanishing reference configurations (e.g., Minkowski or
de Sitter metric, frames, connections) play an essential
role in obtaining these total values.
Our quasilocal expressions likewise
depend on a reference configuration.
A reasonable choice for the reference configuration is to embed
the spatial surface and its boundary into a Minkowski space
\cite{BY93}; more generally one could use (anti) de Sitter space, a
homogeneous cosmology, a Schwarzschild solution, etc.  The evolution
vector field can be selected to correspond to a Killing field
of the reference configuration to obtain the quasilocal
energy-momentum (and angular momentum).  As in Ref. \cite{BY93},
alternate choices  of the evolution vector field can  be used to
distinguish between quasilocal
quantities and conserved charges, mass and energy, etc.

\section{Einstein's theory and a spherically symmetric example}

For Einstein's theory our quasilocal expressions reduce to
only two possibilities depending upon the
choice of boundary conditions---the frame (i.e.,
the metric or intrinsic geometry)
or the connection (the extrinsic geometry)
could be held fixed on the boundary.  These Dirichlet and Neumann
quasilocal expressions are
 \begin{eqnarray}
{\cal B}_{\vartheta} &=&
 i_N {\buildrel \scriptstyle \circ \over \omega}{}^\alpha{}_\beta
   \Delta \epsilon_\alpha{}^\beta
 + \Delta \omega^\alpha{}_\beta
   \wedge i_N \epsilon_\alpha{}^\beta,
 \label{Btheta} \\
 {\cal B}_{\omega}
&=&  i_N \omega^\alpha{}_\beta \Delta \epsilon_\alpha{}^\beta
 + \Delta \omega^\alpha{}_\beta \wedge i_N
 {\buildrel\scriptstyle\circ\over\epsilon}{}_\alpha{}^\beta,
\end{eqnarray}
with the corresponding boundary terms in the Hamiltonian variation
\begin{eqnarray}
 \delta {\cal H}_{\vartheta}(N) &\approx&
 d i_N ( -\Delta \omega^\alpha{}_\beta
   \wedge \delta \epsilon_\alpha{}^\beta ),  \\
 \delta {\cal H}_{\omega}(N) &\approx&
 d i_N ( \delta \omega^\alpha{}_\beta
   \wedge \Delta \epsilon_\alpha{}^\beta ),
\end{eqnarray}
respectively, where $\epsilon^{\alpha\beta}=*(\vartheta^\alpha
\wedge \vartheta^\beta)$.
The expression (\ref{Btheta}) differs slightly from one due to Katz
\cite{Ka85}.  It is also similar to the
famous expression of Brown and York \cite{BY93}.  Significant
differences from the latter are that
(i) our expression is covariant,
(ii) our expression is not specialized to the
timelike boundary being orthogonal to the spacelike hypersurface,
(iii) our expression includes interaction terms between the
physical system and the reference configuration, and
(iv) we have relaxed the
seemingly natural choice {\em lapse\/} $=1$ as
it prevents the quasilocal
energy from attaining to the total energy in the limit for
asymptotically anti-de Sitter solutions.   (There is now a new work
\cite{BCM94}
which also considers quasilocal quantities in
asymptotically non-flat spaces.)

Consider the static spherically symmetric metric
$ d s^2 = -e^{2\Phi} dt^2 + e^{-2\Phi} dr^2
 + r^2 (d \theta^2 + sin^2\theta d \phi^2) $
where $e^{2\Phi}=1-2m/r+\lambda r^2$ and the reference configuration
is given by $m=0$, i.e., Minkowski or (anti) de Sitter.
For the quasilocal energy within a centered sphere we find
$ E_\vartheta = r \alpha( e^{\Phi_0} - e^\Phi)$ and
$E_\omega = r \alpha  e^{\Phi - \Phi_0}(e^{\Phi_0} - e^{\Phi})$.
An appropriate choice of lapse is $\alpha = e^{\Phi_0} =
\sqrt{1+\lambda
r^2}$ which corresponds to the reference configuration timelike
Killing vector.  For this choice both expressions give $m$ for the
total energy, while
the energy within the horizon is $2m$ for $E_\vartheta$ and $\infty$
for $E_\omega$.  The quasilocal energy exterior to the horizon is
negative in both cases.  Note that there is no simple relationship
between the quasilocal energies of Schwarzschild anti-de Sitter space
referenced to Minkowski space, Schwarzschild anti-de Sitter space
referenced to anti-de Sitter space
and de Sitter space referenced to Minkowski space.

\section{Application to black hole thermodynamics}

One application of our expressions is to
black hole thermodynamics \cite{BY93,BCM94,Wa93}.
For this purpose we want to control the quasilocal energy-momentum so
we must allow $N$ to vary.  Hence we Legendre transform to the
``microcanonical'' Hamiltonian
$H_{\mathrm{micro}}(N) := H(N) - \oint_{\partial \Sigma} {\cal B}(N)$.
We choose the connection as one of our control variables, and use
(\ref{inomega}) but drop the unphysical dynamic reference frame
contribution due to
${\hbox{\it \char'44}\!}_N \vartheta^\alpha$.
We take our boundaries
at $\infty$ and on the bifurcate Killing horizon.  For the evolution
vector field $N$ we use the Killing field
$\chi := \partial_t + \Omega_H \partial_\phi$ which is normal to and
vanishes on the horizon. We obtain the ``first law'' for a general
gravity theory by evaluating ``on shell''
\begin{eqnarray}
0 =\delta H_{\mathrm{micro}}(\chi) = \oint_{\partial \Sigma} \delta
{\cal B}(\chi)
  &=&\oint_{\infty} \delta {\cal B}(\chi) - \oint_H \delta {\cal
B}(\chi)   \nonumber \\
  &=&(\delta E + \Omega_H \delta J) -
    \oint_H D_{\alpha} \chi^\beta \delta \rho_{\beta}{}^\alpha.
\end{eqnarray}
The last integral, which for Einstein's theory has
the form $\simeq \kappa \delta A$, identifies the
{\em entropy\/}  for these general geometric gravity theories,

\section{Spinor expressions}

Some of the proposed quasilocal expressions for Einstein's theory
are formulated in terms of auxiliary spinor fields.  Here we
indicate the relationship between such expressions and our expressions
and briefly consider the application of our formalism to spinor
formulations.

Via certain new {\em spinor-curvature\/} identities \cite{NTZ94}
several new {\em quadratic spinor\/} Lagrangians
for Einstein's theory have been found \cite{NT95}.  The different
versions
depend on whether the connection is varied independently and how the
vanishing torsion constraint is imposed.
One of the simplest is
\begin{equation}
{\cal L}_{\mathrm{qs}} \!:=\!
2 D({\overline\psi}\vartheta) \gamma_5 D(\vartheta \psi)
  \!\equiv - {\overline \psi} \psi R \ast\!1
  + d \{ D({\overline \psi} \vartheta) \gamma_5 \vartheta \psi
   \!+\! {\overline \psi} \vartheta \gamma_5 D(\vartheta \psi) \}.
\end{equation}
It differs from the usual Einstein-Hilbert action by just a total
differential. The variables are
a Dirac matrix valued orthonormal frame one-form
$\vartheta := \gamma_\alpha \vartheta^\alpha$
and a ``normalized'' spinor field $\psi$
(i.e., ${\overline \psi} \psi = 1$,
 ${\overline \psi} \gamma_5\psi = 0$).
Asymptotically $\psi \sim \hbox{const} + O(1/r)$ so the Lagrangian
is $O(1/r^4)$ which guarantees finite action.

The corresponding covariant (we have dropped a term proportional to
$i_N\omega$ which generates the frame gauge
transformations and vanishes on shell) Hamiltonian 3-form
has the form
\begin{eqnarray} {\cal H}_{\mathrm{qs}}(N)
&:=& 2\{ D({\overline \psi} {\not\!\! N}) \gamma_5 D(\vartheta \psi)
   + D({\overline \psi}\vartheta) \gamma_5 D({\not\!\! N} \psi) \}
\label{Hqs}\\
 &\equiv& -2 {\overline \psi} \psi N^\mu G_{\mu\nu}
 *\!\vartheta^{\nu}
  + 2d\{ {\overline \psi}{\not\!\!N}\gamma_5 D(\vartheta\psi)
   + D({\overline \psi}\vartheta)\gamma_5 {\not\!\!N}\psi \},
\end{eqnarray}
which is just the ADM Hamiltonian up to a total differential.
It is asymptotically $O(1/r^4)$ and its variation has an $O(1/r^3)$
boundary term which vanishes asymptotically so there is no need
for a further adjustment \cite{RT74} by an additional boundary term.
Expression (\ref{Hqs}) is similar to the Hamiltonian 3-form associated
with the Witten positive energy proof \cite{Ne84}:
\begin{eqnarray}
{\cal H}_{\mathrm{w}}(\psi) &:=& 2 \bigl( D {\overline \psi}\gamma_5
D(\vartheta\psi)
  + D({\overline \psi}\vartheta)\gamma_5 D \psi \bigr) \\
 &\equiv& -2 N^\mu G_{\mu\nu} *\!\vartheta^{\nu}
 + 2d( {\overline \psi} \vartheta \gamma_5 D \psi
  - D {\overline \psi} \gamma_5 \vartheta \psi ), \label{Hwitten}
\end{eqnarray}
wherein $N^{\mu}={\overline \psi} \gamma^{\mu} \psi$.
Again this is the ADM Hamiltonian up to a total differential but
${\cal H}_{\mathrm{w}}$ (unlike ${\cal H}_{\mathrm{qs}}$) is
{\em not\/} related
to a Lagrangian.   For these spinor Hamiltonians once again on a
solution only the boundary terms contribute to the value.

Note that for these spinor expressions there is no explicit need for a
reference configuration, the spinor field {\em implicitly\/} plays
this
role \cite{Lau95}.  In order to compare these spinor expressions with
the quasilocal expressions discussed earlier we introduce an explicit
reference configuration.  We then find that both spinor expressions
are related to our expression (\ref{Btheta}):
\begin{eqnarray}
{\cal B}_{\mathrm{qs}}(N) &:=& 2 \bigl( {\overline \psi}{\not\!\! N}
\gamma_5 D(\vartheta\psi) +
D({\overline \psi} \vartheta) \gamma_5 {\not\!\! N} \psi \bigr) \\
 &=& \Delta \omega^\alpha{}_\beta i_N
\epsilon_\alpha{}^\beta
- 2\bigl({\overline \psi} {\not\!\! N} \gamma_5 \vartheta
{\buildrel \scriptstyle \circ \over D} \psi + {\buildrel \scriptstyle
\circ \over D} {\overline \psi} \gamma_5 \vartheta {\not\!\! N} \psi
\bigr), \\
 {\cal B}_{\mathrm{w}}(\psi) &:=& -2 \bigl( {\overline \psi}\gamma_5
\vartheta D \psi + D {\overline \psi} \gamma_5 \vartheta \psi \bigr)
\label{Bw} \\
 &=& \Delta \omega^\alpha{}_\beta i_N
\epsilon_\alpha{}^\beta - 2 \bigl( {\overline \psi} \gamma_5 \vartheta
{\buildrel \scriptstyle \circ \over D} \psi + {\buildrel \scriptstyle
\circ \over D} {\overline \psi} \gamma_5 \vartheta \psi \bigr).
\end{eqnarray}
In the limit $r \rightarrow \infty$ these spinor
boundary expressions also give the correct total energy-momentum.
Expressions like Eq.  (\ref{Bw}) have been used in several quasilocal
energy investigations \cite{DM91}.

Having introduced a reference configuration, instead of just comparing
the Hamiltonian boundary
expression we can apply our general formalism.  The momenta conjugate
to the spinor fields can be introduced; then, in addition to the
frame and connection type terms we had earlier,
the quasilocal expression acquires spinor terms of the sort
(\ref{Bphi}, \ref{Bp})
and the variation of the Hamiltonian contains extra spinor field
contributions of the type (\ref{symp}).
\section{Discussion}

Which quasilocal expression gives the correct physics?  The physical
role of the spinor field especially still seems mysterious.  One way
to investigate these various quasilocal expressions is to do more
direct calculations for exact solutions, e.g., \cite{Mar94}.
However, a deeper theoretical investigation could be more revealing.
Our formulation provides a good starting point for such an
investigation.
Note that {\em all\/} of the expressions presented here correspond to
the work done in some (ideal) physical process.  The situation is
similar to thermodynamics with its different energies (enthalpy,
Gibbs, Helmholtz, etc.) An even better analogy is the electrostatic
work required while controlling the potential on the boundary of a
region vs.  that required while controlling the charge density
\cite{KT79}.
Thus for the spinor expressions we simply have a different boundary
symplectic structure corresponding to different control variables.
What must be held fixed in each case is found by calculating the
boundary term in the variation of the Hamiltonian.  Mathematically
this is straightforward.  But no matter which technical procedure is
used
for the relation between the frame, the connection and the vanishing
torsion condition the complete results for the variational symplectic
structure turns out to be rather complicated.  Briefly, for the spinor
expressions the main conclusion is, not surprisingly, that we must
hold the orthonormal frame and the spinor field fixed on the boundary.
Hence to understand the physics of the quasilocal spinor expressions
one must understand the physical meaning of controlling the spinor
field on the boundary.  For the Witten Hamiltonian the relation
$N^\mu=\overline\psi\gamma^\mu\psi$ already gives part of the answer.

Thus, for general geometric gravity theories, from a covariant
Hamiltonian formulation using differential forms, by always working
with four dimensionally covariant variables and by using symplectic
ideas, we have found several manifestly covariant Hamiltonian boundary
term expressions for the quasilocal quantities:  energy-momentum and
angular momentum.  Our quasilocal expressions depend only on observer
independent geometric quantities.  They are differential 2-form
expressions which can be evaluated on any closed 2-surface.  These
quasilocal expressions depend on (i) a field configuration, (ii) a
reference configuration (or spinor field), and (iii) an evolution
vector field $N$ on the boundary.  Our formulation should be a good
basis for further investigations aimed at understanding both the
physical meaning of the quasilocal quantities for the different
control modes and the physical interpretation of the contributions
from auxiliary spinor fields.

\section*{Acknowledgment}

We would like to thank V. V. Zhytnikov for his helpful discussions.
This work was supported by the National Science Council of the
Republic of China under contracts NSC 83-0208-M-008-014,
84-2112-M-008-004.

\end{document}